\documentclass[aps,prd,10pt,twocolumn,superscriptaddress,showpacs,footinbib]{revtex4-1}
\pdfoutput=1
\usepackage{hyperref}
\usepackage{slashed}
\usepackage{bm}
\usepackage{graphicx}
\usepackage{enumerate}
\usepackage{amsfonts,amsmath,amssymb,bm,bbm}
\usepackage{color}
\usepackage{epsfig, subfigure}
\usepackage{setspace}
\usepackage{booktabs, tabularx} 
\usepackage{multirow}

\hypersetup{
    pdfnewwindow=true,      
    colorlinks=true,       
    linkcolor=blue,          
    citecolor=blue,        
    filecolor=blue,      
    urlcolor=blue        
}

\newcommand{\be}{\begin{equation}}  
\newcommand{\ee}{\end{equation}}
\newcommand{\bea}{\begin{eqnarray}}  
\newcommand{\eea}{\end{eqnarray}}

\begin{document}

\title{Directional detection of dark matter in universal bound states}

\author{Ranjan Laha}
\affiliation{Kavli Institute for Particle Astrophysics and Cosmology, \\ Department of Physics, Stanford University, Stanford, CA 94035, USA \\ SLAC National Accelerator Laboratory, Menlo Park, CA 94025, USA \\
{\tt rlaha@stanford.edu}\smallskip}

\date{October 12, 2015}
\begin{abstract}
It has been suggested that several small-scale structure anomalies in $\Lambda$CDM cosmology can be solved by strong self-interaction between dark matter particles.  It was shown in Ref.~\cite{Braaten:2013tza} that the presence of a near threshold S-wave resonance can make the scattering cross section at nonrelativistic speeds come close to saturating the unitarity bound.  This can result in the  formation of a stable bound state of two asymmetric dark matter particles (which we call darkonium).  Ref.~\cite{Laha:2013gva} studied the nuclear recoil energy spectrum in dark matter direct detection experiments due to this incident bound state.  Here we study the angular recoil spectrum, and show that it is uniquely determined up to normalization by the S-wave scattering length.  Observing this angular recoil spectrum in a dark matter directional detection experiment will uniquely determine many of the low-energy properties of dark matter independent of the underlying dark matter microphysics.
\end{abstract}

\maketitle

\section{Introduction}
\label{sec:introduction}

It is widely accepted that the Standard Model particles do not make up the entire matter budget of the Universe.  This profound realization that most of the matter in our Universe is not electromagnetically visible comes from observations of the largest to the smallest scales of the Universe~\cite{Planck:2015xua,Steigman:2007xt,Bovy:2012tw,Bhattacharjee:2013exa}.  Solutions to this ``missing light" problem range from a phenomenological modification of Newton's laws~\cite{Famaey:2011kh} to postulating the presence of a new electromagnetically neutral particle, dark matter~\cite{Strigari:2013iaa}.  The latter solution is more appealing and economical as it is able to solve this conundrum at all scales of the Universe.

The search for particle properties of dark matter has been ongoing for several decades~\cite{Gunn:1978gr,Stecker:1978du,Zeldovich:1980st}.  Searches for dark matter are currently pursued in colliders~\cite{Aad:2014vea}, indirect detection~\cite{Danninger:2014xza,Ackermann:2015zua,Dasgupta:2012bd,Laha:2012fg,Ng:2013xha,Murase:2015gea,Ng:2015gfa,Aartsen:2013dxa,Kistler:2009xf,Yuksel:2007dr,Modak:2015uda,Catena:2015iea}, and direct detection~\cite{Akerib:2013tjd,Amole:2015lsj,Cherry:2014wia,Anand:2013yka,Fan:2010gt,Catena:2014epa,Kaplinghat:2013yxa,Panci:2014gga}.

Although the $\Lambda$CDM model of cosmology is fantastically successful at large scales, there are a number of observations at galactic or smaller scales which suggest the incompleteness of this model~\cite{Weinberg:2013aya}.  These small-scale anomalies are typically classified as the missing satellites problem~\cite{Kravtsov:2009gi}, the core vs.~cusp problem~\cite{Walker:2011zu,Oh:2008ww,KuziodeNaray:2007qi} (see Ref.\,\cite{Strigari:2014yea} for exception) and the too big to fail problem~\cite{BoylanKolchin:2011dk,Tollerud:2014zha,Garrison-Kimmel:2014vqa,Papastergis:2014aba,Ferrero:2011au}.  These problems are recognized when one confronts the astrophysical observations with cold dark matter only simulations.

Baryons dominate the scales relevant for these small-scale structure anomalies.  This has driven the interest in using baryons to solve the small-scale structure problems~\cite{Governato:2012fa,Fry:2015rta,Pontzen:2014lma,Zolotov:2012xd,Brooks:2012vi,Brooks:2012ah,Mollitor:2014ara,Chen:2014bea}.

Several particle physics solutions to these problems have also been noted.  Warm dark matter has the potential to solve the missing satellites problem and the too big to fail problem~\cite{Lovell:2013ola,Horiuchi:2013noa}, although there is some dispute~\cite{Schneider:2013wwa}.

A novel solution to the cusp vs core problem and the too big to fail problem is to hypothesize strong self-interaction between dark matter particles first postulated in Ref.~\cite{Spergel:1999mh}.  Since then, a number of models for self-interacting dark matter have been built~\cite{Buckley:2009in,Loeb:2010gj,Tulin:2013teo,Fan:2013tia,Dasgupta:2013zpn,Chu:2014lja,Bringmann:2013vra,Boddy:2014yra,Kouvaris:2014uoa,Wise:2014ola,Petraki:2014uza,Bellazzini:2013foa,Hochberg:2014dra,Wise:2014jva,Kahlhoefer:2013dca,Foot:2014uba,Foot:2014osa,Buckley:2014hja,Wang:2014kja,Cline:2013zca,Ko:2014nha,Cherry:2014xra,Modak:2015npa}.

Given the controversy over these small-scale structure problems, is it possible to determine the dark matter self-interaction cross section independent of the astrophysical data?  The answer is yes, and a model-independent way to describe strong self-interactions between dark matter particles was recently demonstrated in Ref.~\cite{Braaten:2013tza}.  The presence of an S-wave resonance near the scattering threshold of two dark matter particles can lead to enhancements in the nonrelativistic annihilation and self-interaction cross sections.  When the cross section comes close to saturating the unitarity bound, the S-wave scattering length governs the low-energy dynamics of the system~\cite{Braaten:2004rn,Braaten:2003he,Braaten:2007nq,Artoisenet:2010va}.  If the real part of the S-wave scattering length is positive, the resonance is a bound state below the threshold of the two dark matter particles.

If the dark matter particles do not have any annihilation channel, then the bound state of two dark matter particles is stable.  In this limit, the binding energy is determined uniquely by the S-wave scattering length.  We studied the nuclear recoil energy spectrum from a bound-state dark matter scattering in a direct detection experiment in Ref.~\cite{Laha:2013gva}.  We demonstrated that for a certain choice of the dark matter self-interaction cross section, motivated by the small-scale structure anomalies, the bound-state dark matter (which we named darkonium) can break apart during its collision with the nucleus.  The break-up scattering along with the elastic scattering of the darkonium can produce a unique nuclear recoil energy spectrum.

In this work, we study the unique signature in a dark matter directional detection experiment from an incident darkonium.  The observation of this angular recoil spectrum will be a smoking gun signature of the large scattering length in dark matter interactinonrelativisticons.

\section{Directional detection of darkonium}
\label{sec:directional detection of darkonium}

\subsection{Darkonium}

Due to the present excitement about strong dark matter self-interactions at nonrelativistic velocities, we can try to apply the knowledge gained by studying other nonrelativistic systems~\cite{Braaten:2004rn,Braaten:2003he,Braaten:2007nq} to dark matter.    A phenomenological way to explain strong interaction cross sections at nonrelativistic speeds is to postulate the presence of an S-wave resonance near the scattering threshold.  Although this requires fine-tuning, yet this also represents an extremely predictive scenario.

If the underlying parameters of the problem are such that the S-wave scattering length is much larger than the range of interaction between the particles, then the complete dynamics of the system is determined by the S-wave scattering length, which is in general a complex number.  In this case the resultant cross section scales as $1/v^2$, where $v$ is the relative velocity between the two incident particles.  The S-wave scattering length becomes the largest length scale in the problem and hence dictates the dynamics of the system.  The properties of the system become independent of the underlying details of the interaction between the particles and is determined only by the S-wave scattering length.  Such properties are called universal, as any system with a large S-wave scattering length will have the same properties~\cite{Braaten:2004rn}.  If the S-wave scattering length is positive, the resonance is composed of the two incident particles.  Examples of such systems in particle physics are the deuteron (which is a bound state of the neutron and proton), X(3872) (which is a bound state of charm mesons), the diatomic $^4$He molecule and many others.  Numerous examples of such systems exist in the cold atom literature~\cite{Braaten:2004rn}.

The application of this physics to dark matter system makes the resultant properties of dark matter extremely predictive~\cite{Braaten:2013tza}.  The elastic scattering cross section, annihilation cross section, binding energy, lifetime, and structure of the resonance is determined uniquely by the S-wave scattering length.  The annihilation cross section of the dark matter particles and the decay rate of the resonance are proportional to the imaginary part of the S-wave scattering length.  Turning off the annihilation cross section between the dark matter particles automatically makes the resonance stable~\cite{Braaten:2013tza}.  This also implies that the dark matter must be asymmetric in nature.

Denoting the S-wave scattering length as $a$, the self-interaction cross section between two identical  particles of mass $m$ and relative momentum $k$ is 
\begin{eqnarray}
\sigma_{\rm el} = 8\pi a^2/(1+a^2 k^2) \, ,
\label{eq:self-interaction cross section}
\end{eqnarray}
and the binding energy is given by
\begin{eqnarray}
E_B = \dfrac{1}{m a^2} \, .
\label{eq:binding energy}
\end{eqnarray}

\subsection{Directional detection}

Directional detection of dark matter promises a smoking gun signature of the particle properties of dark matter in the solar circle~\cite{Spergel:1987kx}.  Dark matter particles have an isotropic velocity distribution in the Galactic frame, but the motion of the Solar system provides a preferential incident direction of the dark matter particles in the laboratory frame.  This preferential incoming direction of the dark matter particles imprints itself in the angular distribution of the scattered nucleus.  Although the present constraints from directional detection experiments are weak~\cite{Battat:2014van,Vahsen:2011qx,Monroe:2012qma,Miuchi:2010hn,Riffard:2013psa}, it is expected that near future technology can make their sensitivity competitive~\cite{Nygren:2013nda,Gehman:2013mra}.

Dark matter directional detection is the only way to learn about the full dark matter velocity distribution in the solar circle, and this has motivated a number of theoretical studies~\cite{Spergel:1987kx,Gondolo:2002np,Finkbeiner:2009ug,Ahlen:2009ev,Lisanti:2009vy,Bandyopadhyay:2010zj,Billard:2009mf}.  The intrinsically smaller background also implies that a smaller number of events are required to reveal the interactions of dark matter in these detectors.

Directional detection has been studied only for dark matter point particle scattering.  Here we study the directional detection signal due to an incident bound state of dark matter.  There has been an ongoing interest about dark matter bound states~\cite{Kaplan:2009de,Kaplan:2011yj,Wise:2014jva,Detmold:2014qqa,vonHarling:2014kha,Hardy:2014mqa,Krnjaic:2014xza,Hardy:2015boa,Petraki:2015hla}.  In our case, the formation of dark matter bound states is motivated by the hints of strong self-interaction between dark matter particles.  We take a representative value of the dark matter self-interaction cross section, and this determines the S-wave scattering length.  This S-wave scattering length then determines the form factor and the break-up scattering of the bound state during its collision with the nucleus.

The shape of the angular recoil spectrum for a darkonium scattering with a nucleus is completely determined by the S-wave scattering length.  The predictive nature of the underlying physics implies that if we observe a similar angular recoil spectrum in a dark matter directional detection experiment in the future, this will completely determine many low-energy properties of dark matter.  In this case, the value of the S-wave scattering length determined from the angular recoil spectrum will give us information about the binding energy of the bound state and also the self-interaction cross section between the dark matter particles.  The effective theory, which is determined only by the S-wave scattering length, does not fully describe the underlying microphysics of dark matter particle interactions but can be used to compare the predictions from various different models. 

In our work, the overall normalization of the angular spectra is undetermined.  Although we uniquely predict the shape of the angular recoil spectrum, a complete underlying theory will be required to determine the overall normalization in our results.

\subsection{Formalism}

\subsubsection{Elastic scattering of dark matter particle}

The directional detection rate of a dark matter particle is well known and has been discussed extensively in the literature~\cite{Spergel:1987kx,Gondolo:2002np,Finkbeiner:2009ug,Ahlen:2009ev,Lisanti:2009vy,Bandyopadhyay:2010zj,Billard:2009mf}.  We rederive the relevant expressions to introduce the reader to our notation.  

The Feynman diagram of dark matter particle - nucleus elastic scattering can be found in Fig.\,1 of Ref.~\cite{Laha:2013gva}.  The momentum of the incoming and outgoing dark matter particle is denoted by $\bf{P}$ and $\bf{P'}$.  The corresponding kinetic energies are $P^2/2m$ and $P'^2/2m$, respectively, where $m$ denotes the mass of the dark matter particle.   

The momentum of the incoming and outgoing target nucleus is denoted by $\bf{K}$ and $\bf{K'}$ respectively.  Their kinetic energies are denoted by $K^2/2 m_A$ and $K'^2/2 m_A$ respectively, where $m_A$ is the mass of the target nucleus.  We will work in the laboratory frame where the target nucleus is initially at rest and hence ${\bf K}$ = 0.  The momentum transfer is denoted by ${\bf q}$ and ${\bf q} = {\bf K'}$ in the laboratory frame.

The phase space can be written as~\cite{Laha:2013gva}
\begin{eqnarray}
\left( d\Phi \right)_{A+1, \, {\rm Lab}} = \dfrac{q^2 dq \, 2\pi \, d ({\rm cos}\, \theta)}{(2 \pi)^2} \dfrac{m}{q \,P} \, \, \delta \left({\rm cos}\, \theta - \dfrac{q \, m}{2 \mu \, P}\right) \, .\phantom{1111}
\label{eq:nonrelativistic 2-body phase space simplified}
\end{eqnarray}
The angle $\theta$ is defined by the dot product ${\bf P}{\bf .}{\bf q} = P q \,{\rm cos} \, \theta$.  The reduced mass of the dark matter particle and the nucleus is denoted by $\mu$.

We denote the matrix element of the elastic scattering by $- i \, G_A (q)$ where the exact form of $G_A (q)$ is determined by the microphysics of scattering between the dark matter particle and the nucleus.  The normalization of $G_A(q)$ is an arbitrary constant in this work.  Although we will take a specific form of $G_A(q)$ while showing our results, we remind the reader that the normalization of all our results is arbitrary.  The S-wave scattering length uniquely determines the shape of the recoil spectrum but does not say anything about its normalization. 

The differential nuclear recoil energy is given by $dE_{\rm nr} = q \,dq/m_A$.  The expression for $d(\sigma v)$ is given by $|G_A(q)|^2 \, \left( d\Phi \right)_{A+1,{\rm lab}}$.  From this expression, we derive
\begin{eqnarray}
\left(\dfrac{d(\sigma v)}{dE_{\rm nr}}\right)_{A+1} &=& \dfrac{|G_A(q)|^2}{q \, dq} m_A \dfrac{q^2 \, dq \, 2 \pi \, d({\rm cos} \, \theta)}{(2 \pi)^2} \dfrac{m}{q \, P} \nonumber\\
&\times& \delta \left({\rm cos} \, \theta - \dfrac{q \, m}{2 \mu \, P}\right) \,.
\label{eq:particle elastic scattering mid-step1}
\end{eqnarray}
The following double differential is easily derived from Eqn.~\ref{eq:particle elastic scattering mid-step1}:
\begin{eqnarray}
\left(\dfrac{d^2(\sigma v)}{dE_{\rm nr} \, d\Omega}\right)_{A+1} &=& \dfrac{|G_A(q)|^2 \, m_A \, m}{4 \pi^2 \, P}
\, \delta \left({\rm cos} \, \theta - \dfrac{q \,m}{2 \mu \,P}\right) \,.
\label{eq:particle elastic scattering no f(v)}
\end{eqnarray}
In the above expression, the differential solid angle is given by $d\Omega$ = 2$\pi$ $d$(cos\,$\theta)$.  We numerically checked that when we integrate this expression over the solid angle, we reproduce the relevant expression in Ref.~\cite{Laha:2013gva}.

Since present directional dark matter detectors like DRIFT~\cite{Battat:2014van} are primarily sensitive to spin-dependent scattering, we can compare Eqn.~\ref{eq:particle elastic scattering no f(v)} to the standard expression used in the literature~\cite{Peter:2013aha} to obtain 
\begin{eqnarray}
|G_A (q)|^2 = \dfrac{\pi}{\mu^2} \, \sigma^{\rm SD}_A \, F^2_{\rm SD} (E_{\rm nr}) \, ,
\label{eq:expression for |G_A(q)|^2}
\end{eqnarray}
where $\sigma^{SD}_A$ refers to the spin-dependent cross section between the dark matter particle and nuclei.  In the expression for this cross section, it is convenient to include a multiplicative factor of the spin-dependent cross section between the dark matter particle and the proton~\cite{Peter:2013aha}.

Since the dark matter particles in our galaxy have a normalized velocity distribution, denoted by $f(v)$, the interaction rate of elastic scattering between dark matter particles and nuclei in the Galactic frame at the solar radius is given by
\begin{eqnarray}
\left(\dfrac{d^2 R}{dE_{\rm nr} \, d\Omega} \right)_{A+1,\, {\rm Gal}} &&= N_T \, n_\chi \int d^3 {\bf v} \, f(v) \nonumber\\
&\times& \dfrac{\sigma^{\rm SD}_A \, F^2_{\rm SD} (E_{\rm nr}) \, m_A}{4\pi \, \mu^2} \delta \left({\bf v}.\hat{q} - \dfrac{q }{2 \mu }\right) \, ,\phantom{111}
\label{eq:dark matter elastic Galactic frame}
\end{eqnarray}
where $\hat{q}$ represents the unit vector in the direction of $\bf q$.  From the definition of cos\,$\theta$, we find that ${\bf v}.\hat{q} = v \, {\rm cos}\,\theta$.  Here we denote the speed of the dark matter particle in the Galactic frame by $v$.  The local number density of dark matter particles is denoted by $n_\chi$, and $N_T$ represents the number of the target nuclei.

To obtain the interaction rate in the laboratory frame, we need to boost this expression to the laboratory frame using Galilean kinematics, since all the velocities involved are $\sim \mathcal{O}(100$ km s$^{-1})$.  This is most easily demonstrated by the use of the Radon transform~\cite{Gondolo:2002np}.

Instead of using the Radon transform, which is suitable only for elastic scattering between two particles, we boost our expression to the Galactic frame by a change of coordinates~\cite{Gondolo:2002np}.  This method of boosting the expression from the Galactic frame to the laboratory frame turns out to be especially convenient when we consider darkonium break-up scattering.  

Given that the velocity of the dark matter in the Galactic frame is $\bf v$, the velocity of the dark matter in the laboratory frame is ${\bf v'} = {\bf v} - {\bf v_E}$, where ${\bf v_E}$ is the velocity of the Earth with respect to the Galaxy.  Since the particle number in a differential velocity volume element is conserved, we have $f(v) \, d^3{\bf v} = f'(v') \, d^3{\bf v'}$ where $f'(v')$ is the dark matter velocity distribution in the laboratory frame. 

In the laboratory frame, the velocity-dependent part in Eqn.~\ref{eq:dark matter elastic Galactic frame} reads as $\int d^3 {\bf v'} \, f' (v') \, \delta ({\bf v'}.\hat{q} - q/2 \mu)$.  Using the conservation of the particle number in a differential velocity volume element and inserting the expression of ${\bf v}'$, the break-up scattering in Eqn.~\ref{eq:dark matter elastic Galactic frame} becomes $\int 2\pi \, v^2 \,dv \, d\, {\rm cos}\, \theta_{vq} \, f(v) \, \delta (v \, {\rm cos} \, \theta_{vq} - v_E \, {\rm cos} \, \theta_{v_{E} q} - q/2\mu)$.  Here we define the angles $\theta_{vq}$ and $\theta_{v_E q}$ by the following dot products: ${\bf v}.\hat{q} = v \,{\rm cos} \, \theta_{vq}$ and ${\bf v_E}.\hat{q} = v_E \,{\rm cos} \, \theta_{v_E q}$.  The argument in the delta function also gives the minimum dark matter speed required to cause a recoil of momentum $q$ and in the angle $\theta_{v_E q}$: $v \geq v_E \, {\rm cos}\, \theta_{v_E q} + q/2\mu \equiv v_{\rm min}$.

Integrating over the angle $\theta_{vq}$, we get 
\begin{eqnarray}
\left(\dfrac{d^2 R}{dE_{\rm nr} \, d\Omega_{v_E q}} \right)_{A+1,\, {\rm Gal}} &&= N_T \, n_\chi \int _{v_{\rm min}} ^{v_{\rm max}} \dfrac{\sigma^{\rm SD}_A \, F^2_{\rm SD} (E_{\rm nr}) \, m_A}{4\pi \, \mu^2} \nonumber\\
&\times& 2\pi \,v \, f(v) \, dv \, ,
\label{eq:eq:dark matter elastic Galactic frame integrated over theta_vq}
\end{eqnarray}
where $v_{\rm max}$ represents the maximum dark matter speed in the solar radius.  Here $d\Omega_{v_E q} = 2\pi \, d({\rm cos}\,\theta_{v_E q})$ represents the solid angle that can be measured in a directional detection experiment in the laboratory.

Similar to our previous paper~\cite{Laha:2013gva}, we take 
\begin{eqnarray}
f(v) &=&N\, {\rm exp}(-v^2/2 v_0^2) \, \Theta(v_{\rm max}-v) \,,
\label{eq:f(v)}
\end{eqnarray}
and
\begin{eqnarray}
N = \dfrac{1}{4\pi \left\{-v_0^2 \, v_{\rm max} \, e^{-\dfrac{v_{\rm max}^2}{2 v_0^2}} + \sqrt{\dfrac{\pi}{2}} v_0^3 \, {\rm erf}\left(\dfrac{v_{\rm max}}{\sqrt{2}v_0} \right)\right\}} \, ,  \phantom{11111}
\label{eq:normalization in f(v)}
\end{eqnarray}
where the following values are taken as constants: $v_E$ = 242 km s$^{-1}$, $v_{\rm max}$ = 600 km s$^{-1}$, and $v_0$ = 230 km s$^{-1}$.  The normalization constant $N$ is obtained from $\int d^3 v \, f(v) = 1$.  We neglect the rotational motion of the Earth and the motion of the Earth around the Sun for simplicity~\cite{Lee:2012pf}.  Although using a different dark matter velocity distribution can produce a different recoil distribution~\cite{Chaudhury:2010hj,Kuhlen:2009vh}, and observations~\cite{Bhattacharjee:2012xm,Bhattacharjee:2013exa} and simulations~\cite{Mao:2012hf,Mao:2013nda} do indeed show a non-Maxwellian behavior of the dark matter velocity profile, our choice is dictated by simplicity and intended as a proof of concept.  It is difficult to obtain an analytical form for the double differential while using a non-Maxwellian velocity distribution.  In Sec.~\ref{sec:non-Maxwellian}, we will compare the angular nuclear recoil spectrum due to a Maxwellian distribution to that due to a Tsallis distribution.

The velocity integral can be done analytically and we obtain (in units of GeV$^{-1}$ s$^{-1}$ sr$^{-1}$)
\begin{eqnarray}
&&\left(\dfrac{d^2 R}{dE_{\rm nr} \, d\Omega_{v_E q}} \right)_{A+1,\, {\rm Lab}} = N_T \, n_\chi  \dfrac{\sigma^{\rm SD}_A \, F^2_{\rm SD} (E_{\rm nr}) \, m_A}{4\pi \, \mu^2} \nonumber\\
&\times&  2\pi N\, v_0^2 \left(e^{-\dfrac{v_{\rm min}^2}{2v_0^2}} - e^{-\dfrac{v_{\rm max}^2}{2 v_0^2}} \right) \,, 
\label{eq:dark matter elastic Galactic frame final expression}
\end{eqnarray}
as the full expression for the double differential elastic scattering rate for a dark matter particle with the target in the laboratory.  This expression shows that the dependence on the angle $\theta_{v_E q}$ comes from the exponential term in the second line of Eqn.~\ref{eq:dark matter elastic Galactic frame final expression}.

\subsubsection{Elastic scattering of darkonium}
The analytical expression for the rate of elastic scattering of darkonium is very similar to the expression for the elastic scattering of dark matter particles as detailed in the previous subsection.  We will assume that the local dark matter density is fully composed of darkonium.  In such a case, the number density of incident darkonium is denoted by $n_{\chi_2}$.  The expressions for the scattering rate include the form factor of the darkonium which naturally arises from the calculation~\cite{Laha:2013gva}.

The Feynman diagram of the darkonium - nucleus elastic scattering can be found in Fig. 2 of Ref.~\cite{Laha:2013gva}.  The momentum of the incoming and outgoing darkonium is denoted by ${\bf P}$ and ${\bf P'}$ respectively.  The corresponding energies are given by $-E_B + P^2/4m$ and $-E_B + P'^2/4m$.  The momentum and kinetic energies of the target and scattered nucleus have the same notation as for elastic scattering of dark matter particles.  The momentum transferred in the laboratory frame is denoted by ${\bf q}$.

The phase space for this scattering can be written as 
\begin{eqnarray}
\left( d\Phi \right)_{A+2, \, {\rm Lab}} = \dfrac{q^2 dq \, d ({\rm cos}\, \theta)}{\pi} \dfrac{2m}{2 q \, P} \, \delta \left({\rm cos}\, \theta - \dfrac{q \,}{2 \mu_2 \,} \dfrac{2m}{P}\right) \, ,\phantom{1111}
\label{eq:nonrelativistic darkonium elastic phase space simplified}
\end{eqnarray}
where the angle $\theta$ represents the angle between the incoming darkonium momentum and the momentum transferred in the elastic collision.  The reduced mass of the darkonium - nucleus system is denoted by $\mu_2$.  This expression for the phase space differs from that in Eqn.~\ref{eq:nonrelativistic 2-body phase space simplified} by the presence of $\mu_2$ instead of $\mu$.

The matrix element for this process is~\cite{Laha:2013gva}
\begin{eqnarray}
\mathcal{M} = - G_A(q) \, \dfrac{8 \gamma}{q} \, {\rm tan}^{-1} \dfrac{q}{4 \gamma} \, ,
\label{eq:matrix element darkonium elastic}
\end{eqnarray}
where $\gamma$ denotes the inverse of the S-wave scattering length.

Combining Eqns.~\ref{eq:nonrelativistic darkonium elastic phase space simplified} and \ref{eq:matrix element darkonium elastic}, we get 
\begin{eqnarray}
&&\left(\dfrac{d^2 R}{dE_{\rm nr} \, d\Omega}\right)_{A+2, \, {\rm Gal}} = N_T \, n_{\chi_2} \int d^3{\bf v}\, f(v) \, \dfrac{\sigma_A^{\rm SD} \, F_{\rm SD}^2 (E_{\rm nr})}{\mu^2} \nonumber\\
&\times& \left(\dfrac{4\gamma}{q} \right)^2 \, \left({\rm tan}^{-1} \dfrac{q}{4\gamma} \right)^2 \, \dfrac{m_A}{\pi} \, \delta \left({\bf v}.\hat{q} - \dfrac{q}{2 \mu_2}\right) \, .
\label{eq:differential sigmav darkonium elastic scattering}
\end{eqnarray}
Numerically integrating the expression in Eqn.~\ref{eq:differential sigmav darkonium elastic scattering} over the solid angle reproduces the relevant expression in Ref.~\cite{Laha:2013gva}.  
 
 This expression is very similar to the expression in Eqn.~\ref{eq:dark matter elastic Galactic frame} and this is expected as in both cases we have elastic scattering between two objects.  The difference in this expression is the appearance of $n_{\chi_2}$, the dependence of the delta function on $\mu_2$, the presence of the darkonium form factor $(4\gamma/q)$\,${\rm tan}^{-1} (q/4\gamma)$ and an overall factor of 4.  The factor of 4 can be understood as the coherence factor of the darkonium which is composed of 2 dark matter particles.

To boost this expression to the laboratory frame, we follow the same procedure as given in the previous subsection.  For our choice of the velocity profile, the integration over the velocity can be done analytically and we obtain
\begin{eqnarray}
&&\left(\dfrac{d^2 R}{dE_{\rm nr} \, d\Omega_{v_E q}} \right)_{A+2,\, {\rm Lab}} = N_T \, n_{\chi_2}  \dfrac{\sigma^{\rm SD}_A \, F^2_{\rm SD} (E_{\rm nr}) \, m_A}{\pi \, \mu^2} \nonumber\\
&\times&  2\pi \, N
\left(\dfrac{4\gamma}{q} \right)^2 \, \left({\rm tan}^{-1} \dfrac{q}{4\gamma} \right)^2 
 v_0^2 \nonumber\\
 &\times& \left(e^{-\dfrac{v_{\rm min_2}^2}{2v_0^2}} - e^{-\dfrac{v_{\rm max}^2}{2 v_0^2}} \right) , \phantom{111}
\label{eq:darkonium elastic Galactic frame final expression}
\end{eqnarray}
where  $v_{\rm min_2} \equiv v_E \, {\rm cos}\, \theta_{v_E q} + q/2\mu_2$.  This gives the full expression for the double differential elastic scattering rate for a darkonium with the target in the laboratory.

\subsubsection{Break up scattering of darkonium}
When the binding energy of the darkonium is low enough, it can break up into its constituents during its scattering with the target nucleus.  Here we detail our calculation of this break-up scattering.

The Feynman diagram of the darkonium - nucleus break-up scattering can be found in Fig. 3 of Ref.~\cite{Laha:2013gva}.  The incoming momentum and energy of the darkonium are given by ${\bf P}$ and $-E_B+P^2/4m$.  The two dark matter particles in the final state have a momentum and energy of ${\bf p_1}$, ${\bf p_2}$ and $p_1^2/2m$, $p_2^2/2m$ respectively.  The momentum and energy of the target and scattered nucleus are the same as given in the previous subsection.

The phase space of this configuration is given by~\cite{Laha:2013gva}
\begin{eqnarray}
&&(d \Phi)_{A+1+1, \, {\rm Lab}} = \dfrac{d^3 {\bf q}}{(2\pi)^3} \dfrac{d^3 {\bf r}}{(2\pi)^3} \nonumber\\
&\times& 2\pi \, \delta \left(\dfrac{{\bf P}.{\bf q} - 2 r^2}{2m} - E_B - \dfrac{q^2}{2 \mu_2}\right) \, ,
\label{eq:darkonium break up phase space simplified}
\end{eqnarray}
where ${\bf r} = ({\bf p_1}-{\bf p_2})/2$.  

The matrix element for the three diagrams is given in Ref.~\cite{Laha:2013gva}.  Multiplying the square of the matrix element with the phase space, we obtain
\begin{eqnarray}
&&d(\sigma v) = \dfrac{d^3 {\bf r}}{(2\pi)^3} \dfrac{d^3 {\bf q}}{(2\pi)^3} 2\pi \, \delta  \left(\dfrac{{\bf P}.{\bf q} - 2 r^2}{2m} - E_B - \dfrac{q^2}{2 \mu_2}\right) \nonumber\\
&\times& 16 m^2 \, \dfrac{16 \pi \gamma}{m^2} \, |G_A(q)|^2 \Bigg| \dfrac{1}{4\gamma^2 + (2{\bf r}-{\bf q})^2} + \dfrac{1}{4\gamma^2 + (2{\bf r}+{\bf q})^2} \nonumber\\
&-& \dfrac{i}{2 q (\gamma + i r)} \, {\rm ln} \dfrac{4r^2 + (2\gamma - i q)^2}{4 \gamma^2 + (q - 2r)^2}\Bigg|^2 \, .
\label{eq:d(sigma v) for darkonium break up}
\end{eqnarray}
To arrive at a closed form expression for the scattering rate of darkonium break up interaction, we need to integrate over ${\bf r}$.  This is most conveniently calculated by thinking of the whole-squared term in Eqn.~\ref{eq:d(sigma v) for darkonium break up} as the sum of two terms where we consider the first two terms together as one term and then consider the third term separately.

We first consider the integral over ${\bf r}$ for the first two terms in Eqn.~\ref{eq:d(sigma v) for darkonium break up}:
\begin{eqnarray}
&&\int \dfrac{r^2 dr \, {\rm cos}\, \theta_{qr}}{(2\pi)^2} \, 2\pi \, \delta  \left(\dfrac{{\bf P}.{\bf q} - 2 r^2}{2m} - E_B - \dfrac{q^2}{2 \mu_2}\right) \nonumber\\
&\times& 16 m^2 \, \dfrac{16 \pi \gamma}{m^2} \, |G_A(q)|^2 \Bigg| \dfrac{1}{4\gamma^2 + q^2 + 4r^2 - 4 q r \, {\rm cos} \, \theta_{qr}} \nonumber\\
&+& \dfrac{1}{4\gamma^2 + q^2 + 4r^2 + 4 q r \, {\rm cos} \, \theta_{qr}} \Bigg|^2 \, .
\label{eq:integral over r first two terms}
\end{eqnarray}
The integral over the angle $\theta_{qr}$ is easily calculated.  The delta function is rewritten as $m/(2 r_{\theta_{Pq}}) \, \delta (r - r_{\theta_{Pq}})$, where $r_{\theta_{Pq}}^2 = - \gamma^2 + {\bf P}.{\bf q}/2 - (m q^2)/(2 \mu_2)$.  Requiring $r_{\theta_{Pq}} \geq 0$, we get the threshold condition for a given angle $\theta_{Pq}$: $\theta_{Pq} \geq {\rm cos}^{-1} \, (m E_B + m q^2/2 \mu_2)/(m v q)$.  The integration over the variable $r$ is now accomplished using the delta function, where the maximum value of $r$ is $R = \sqrt{m q v - \gamma^2 - (m q^2)/2\mu_2}$.

The integration over the remaining terms in Eqn.~\ref{eq:d(sigma v) for darkonium break up} is evaluated in a similar manner.  We then arrive at the following analytical expression for the double differential:
\begin{widetext}
\begin{eqnarray}
&&\dfrac{d^2 (\sigma v)}{dE_{\rm nr} \, d\Omega} = \dfrac{m_A \, q}{(2\pi)^3} \, 16 m^2 \, \dfrac{16 \pi \gamma}{m^2} |G_A (q)|^2 
\, \Theta \left(v - \dfrac{\left(\dfrac{\gamma^2}{m q} + \dfrac{q}{2 \mu_2}\right)}{{\rm cos}\, \theta_{Pq}} \right)  \dfrac{m r_{\theta_{Pq}}}{4\pi} (A\,+\,B\,+\,C) \, , \phantom{11}
\label{eq:double differential sigma v darkonium break up}
\end{eqnarray}
where
\begin{eqnarray}
&&A = \dfrac{4}{(4 \gamma^2 + q^2 + 4 r_{\theta_{Pq}}^2)^2} 
 \Bigg( \dfrac{1}{1-\dfrac{16 q^2 r_{\theta_{Pq}}^2}{(4 \gamma^2 + q^2 + 4 r_{\theta_{Pq}}^2)^2}} + 
\dfrac{4 \gamma^2 + q^2 + 4 r_{\theta_{Pq}}^2}{4 q r_{\theta_{Pq}}} \, {\rm tanh}^{-1} \dfrac{4 q r_{\theta_{Pq}}}{4 \gamma^2 + q^2 + 4 r_{\theta_{Pq}}^2} \Bigg)\, ,
\label{eq:A}
\end{eqnarray}
\begin{eqnarray}
B= 2\,  \Bigg| \dfrac{i}{2q (\gamma + i r)} \, {\rm ln} \dfrac{4 r_{\theta_{Pq}}^2 + (2 \gamma - i q)^2}{4 \gamma^2 + (q - 2 r_{\theta_{Pq}})^2} \Bigg|^2 \, ,
\label{eq:B}
\end{eqnarray}
and
\begin{eqnarray}
&&C = \dfrac{1}{q r_{\theta_{Pq}}} \, {\rm tanh}^{-1} \left(\dfrac{4 q r_{\theta_{Pq}}}{4 \gamma^2 + q^2 + 4 r_{\theta_{Pq}}^2} \right) 
 \Bigg( - \dfrac{i}{2 q (\gamma + i r_{\theta_{Pq}})} \, {\rm ln} \dfrac{4 r_{\theta_{Pq}}^2 + (2 \gamma - i q)^2}{4 \gamma^2 + ( q - 2 r_{\theta_{Pq}})^2} \nonumber\\
&+& \left\{ - \dfrac{i}{2 q (\gamma + i r_{\theta_{Pq}})}  \, {\rm ln} \dfrac{4 r_{\theta_{Pq}}^2 + (2 \gamma - i q)^2}{4 \gamma^2 + ( q - 2 r_{\theta_{Pq}})^2} \right\}^* \Bigg) \, .
\label{eq:C}
\end{eqnarray}
\end{widetext}
 Numerically integrating the expression in Eqn.~\ref{eq:double differential sigma v darkonium break up} over the solid angle reproduces the relevant expression in Ref.~\cite{Laha:2013gva}.
  
We multiply Eqn.~\ref{eq:double differential sigma v darkonium break up} by $N_T \, n_{\chi_2} \, f(v)$ and integrate over the velocity volume element $d^3 {\bf v}$ to obtain the double differential scattering rate for darkonium break up in the Galactic frame.

To boost the expression to the laboratory frame, we again follow the change of co-ordinates strategy as outlined previously.  Using the change of variable ${\bf v} \rightarrow {\bf v'} = {\bf v} - {\bf v_E}$, we have
\begin{eqnarray}
r_{\theta_{Pq}}^2 \rightarrow \tilde{r}_{\theta_{Pq}}^2 = m ({\bf v}-{\bf v_E}).{\bf q} - \gamma^2 - \dfrac{m q^2}{2\mu_2} \, .
\label{eq:rtilde}
\end{eqnarray}

The change of variable for the theta-function in Eqn.~\ref{eq:double differential sigma v darkonium break up} is accomplished by a change in the velocity co-ordinate in the definition of $r_{\theta_{Pq}}$ and demanding the resulting expression to be greater than zero.  This gives us the minimum Galactic dark matter velocity required to break up a darkonium and have the nucleus scattered in the angle $\theta_{v_E q}$:
\begin{eqnarray}
v \geq \dfrac{1}{{\rm cos}\, \theta_{vq}} \left(\dfrac{\gamma^2}{m q} + \dfrac{q}{2 \mu_2} + v_E \, {\rm cos} \theta_{v_E q}\right) \, .
\label{eq:vmin for darkonium break up}
\end{eqnarray}

We get the following analytical expression for the double differential scattering rate of darkonium break up in the laboratory:
\begin{eqnarray}
&&\left(\dfrac{d^2 R}{dE_{\rm nr} \, d\Omega_{v_E q}} \right)_{A+1+1, \, {\rm Lab}} = N_T \, n_{\chi_2} \nonumber\\
&\times& \int v^2 \, dv \, d\Omega_{vq} f(v) \dfrac{m_A q}{(2\pi)^3} \, 16 m^2 \, \dfrac{16 \pi \gamma}{m^2} |G_A(q)|^2 \nonumber\\
&\times& \Theta \left(v - \dfrac{\dfrac{\gamma^2}{m q} + \dfrac{q}{2 \mu_2} + v_E \, {\rm cos}\, \theta_{v_E q}}{{\rm cos}\, \theta_{vq}} \right)  \dfrac{m r_{\theta_{vq}}}{4\pi} \nonumber\\
&\times& (\tilde{A}\,+\, \tilde{B}\,+\, \tilde{C}) \, ,
\label{eq:darkonium break up scattering rate in the laboratory}
\end{eqnarray}
where $\tilde{A}$, $\tilde{B}$, and $\tilde{C}$ are the expressions in Eqns.~\ref{eq:A}, \ref{eq:B}, and \ref{eq:C} with $\tilde{r}_{\theta_{Pq}}$ replacing $r_{\theta_{Pq}}$.  We performed the integration over the velocity volume element numerically.

Having the complete expressions for the dark matter elastic scattering, darkonium elastic scattering, and darkonium break-up scattering in Eqns.~\ref{eq:dark matter elastic Galactic frame final expression}, \ref{eq:darkonium elastic Galactic frame final expression}, and \ref{eq:darkonium break up scattering rate in the laboratory} respectively, we proceed to calculate the angular dependence of these interactions when considering the nuclear recoil energy over certain energy bin.

\subsection{Results}

We first consider the recoil angular distributions when we take the dark matter particle mass $m = 100$ GeV, and $\sigma_{\rm el}/m$ = 1 cm$^2$ g$^{-1}$ at relative velocity $v = 10$ km s$^{-1}$.  In this case, the darkonium binding energy is 0.52 keV, and it breaks apart during its collision with the nucleus.  The nuclear recoil spectrum in this case was shown in Ref.~\cite{Laha:2013gva}.

As a variation, we also consider the case when the dark matter particle mass is 10 GeV and $\sigma_{\rm el}/m$ = 1 cm$^2$ g$^{-1}$ at $v = 10$ km s$^{-1}$.  The darkonium binding energy is 52 keV in this case and it does not break apart during its collision with the nucleus.  The nuclear recoil spectrum in this case was also shown in Ref.~\cite{Laha:2013gva}. 

The targets used in the directional detection experiment typically involve $^{19}{\rm F}$~\cite{Battat:2014van,Vahsen:2011qx,Monroe:2012qma,Miuchi:2010hn,Riffard:2013psa}.  Recently there has been an interest in using xenon as a target in directional detection experiments~\cite{Nygren:2013nda,Gehman:2013mra,Mohlabeng:2015efa,Li:2015zga}.  We show our results for $^{19}{\rm F}$ and ${\rm Xe}$ targets.  We only choose the isotopes of xenon, $^{129}{\rm Xe}$ and $^{131}{\rm Xe}$, which are sensitive to spin-dependent interactions~\cite{Aprile:2013doa}.

The normalization in all our plots is arbitrary and is not governed by the S-wave scattering length.  For concreteness, we take $\sigma_{\rm SD}^p$ = 10$^{-39}$ cm$^2$.  We take the details of the form factors from Refs.~\cite{Aprile:2013doa,Pato:2011de,Bednyakov:2006ux,Ressell:1997kx}.  For $^{129}{\rm Xe}$ and $^{131}{\rm Xe}$, we take the Bonn A coefficients from Ref.~\cite{Ressell:1997kx}.  We take $a_0 = 0$ and $a_1 = 2$ in the definition of the spin-dependent form factors for all cases~\cite{Peter:2013aha}.

The local dark matter density is taken to be 0.3 GeV cm$^{-3}$.  While showing our results for the dark matter particle and darkonium, we will assume that the full local density of dark matter is composed of individual dark matter particles and darkonium respectively.  Since the darkonium has double the mass of the dark matter particle, the number density of dark matter particles is double that of the darkonium.

\begin{figure*}[!thpb]
\centering
\includegraphics[angle=0.0,width=0.43\textwidth]{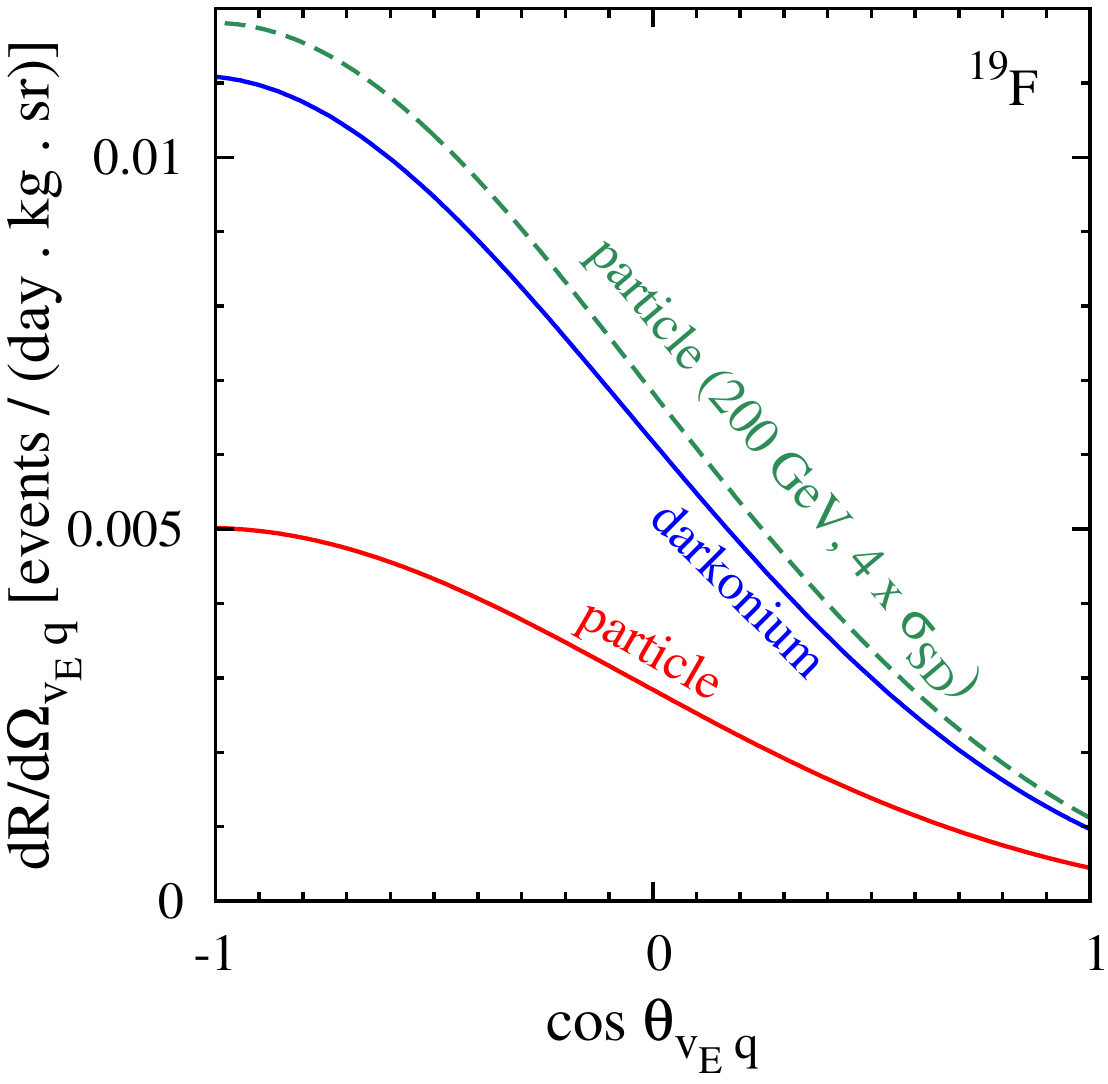}
\includegraphics[angle=0.0,width=0.43\textwidth]{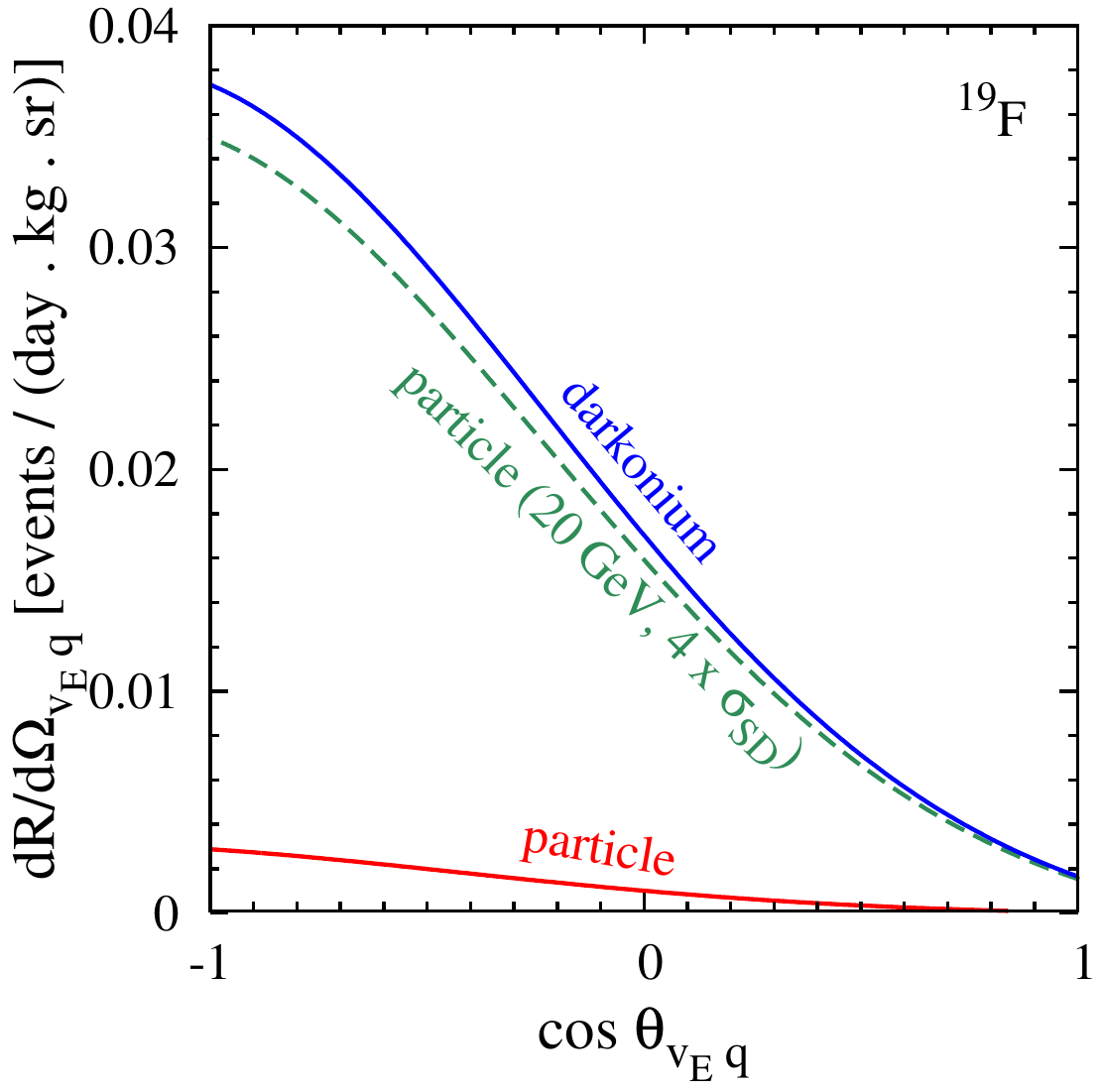}
\caption{The angular nuclear recoil spectra for dark matter particle (of mass $m$) scattering (red),  darkonium scattering (blue),  and for a dark matter particle with mass $2m$ and $\sigma_{\rm SD}$ that is 4 times larger (dashed green) with $^{19}$F as the target.  We have determined the S-wave scattering length, which uniquely determines the shape of the angular recoil spectrum, by taking the elastic scattering cross section per unit mass to be $\sigma_{\rm el}/m$ = 1 cm$^2$ g$^{-1}$ at relative velocity $v = 10$ km s$^{-1}$.   The normalizations of the curves correspond to the choice
$\sigma^p_{SD} = 10^{-39}~{\rm cm}^2$.  {\bf Left plot :} The dark matter particle mass is taken to be 100 GeV and the energy bin for integration is [5, 40] keV.  {\bf Right plot :} The dark matter particle mass is taken to be 10 GeV and the energy bin for integration is [5, 14] keV. Note the different scales in the y-axis.}
\label{fig:Angular spectra 1}
\end{figure*}

\begin{figure*}[!thpb]
\centering
\includegraphics[angle=0.0,width=0.43\textwidth]{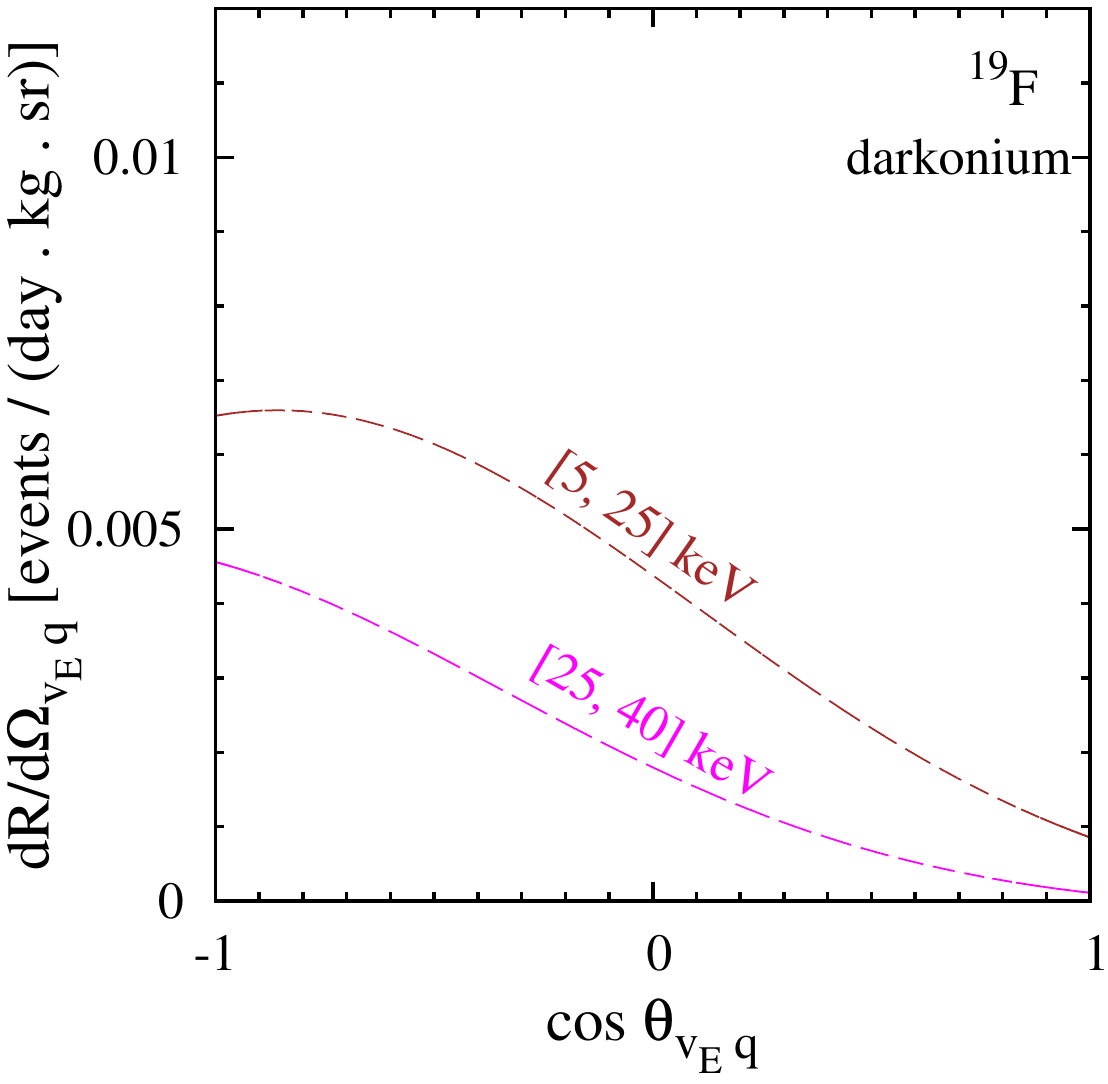}
\includegraphics[angle=0.0,width=0.43\textwidth]{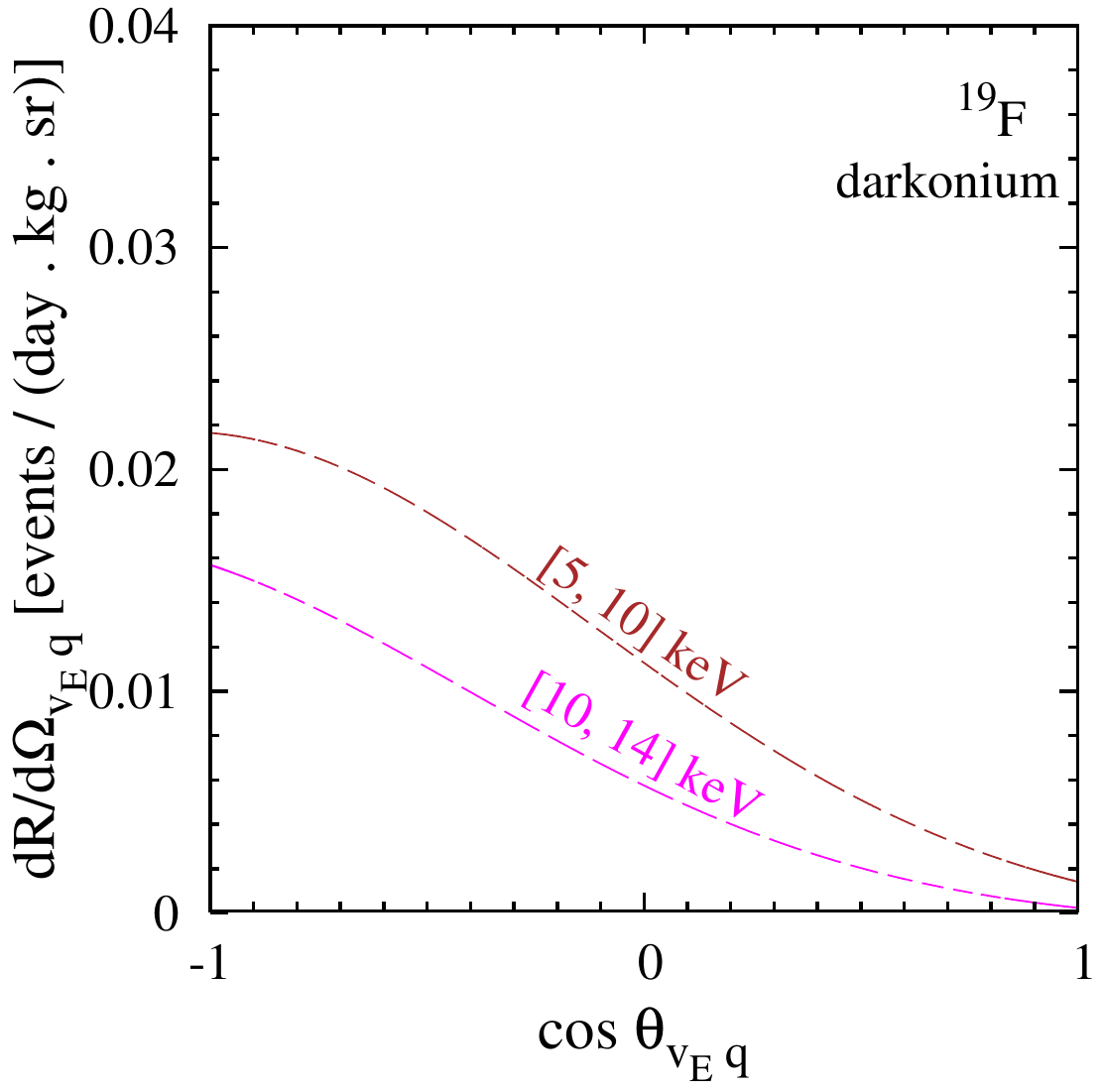}
\caption{The angular recoil spectra for darkonium scattering with the nucleus when different energy bins are considered.  We have taken $^{19}$F as the target.  {\bf Left plot :} The darkonium is composed of two dark matter particles each of which is 100 GeV.  The angular recoil spectra in the [5, 25] keV and [25, 40] keV energy bins are shown by brown and magenta dashed lines, respectively.  {\bf Right plot:}  The darkonium is composed of two dark matter particles, each of which is 10 GeV.  The angular recoil spectra in the [5, 10] keV and [10, 14] keV energy bins are shown by brown and magenta dashed lines, respectively.  Note the different scales in the y-axis.}
\label{fig:Angular spectra energy bin 1}
\end{figure*}

\subsubsection{Target: Fluorine}
We show the angular recoil distribution of nuclear scattering for 100 GeV dark matter and $\sigma_{\rm el}/m$ = 1 cm$^2$ g$^{-1}$ at $v = 10$ km s$^{-1}$ on the left in Fig.~\ref{fig:Angular spectra 1}.  The target nucleus is $^{19}{\rm F}$.  We have integrated over the energy bin 5 keV to 40 keV to obtain the angular recoil spectra for the 100 GeV and 200 GeV dark matter particle scattering, and darkonium scattering.  The threshold used in this calculation follows from Ref.~\cite{Grothaus:2014hja}.  For the case of 200 GeV particle scattering, we have taken the cross section to be 4 times larger, i.e., $\sigma_{\rm SD}^p$ = 4 $\times$ 10$^{-39}$ cm$^2$, as this produces a comparable recoil energy spectrum~\cite{Laha:2013gva}.  

The angular spectrum for the darkonium scattering is very different compared to the 100 GeV particle scattering both in shape and normalization.  There is similarity with the angular spectrum from the 200 GeV particle scattering but even here the shapes are different.  The width between the angular spectra of darkonium scattering and the 200 GeV particle scattering varies with angle and this will be an important experimental discriminator.  The number of events in different angular bins for a 200 GeV dark matter particle and darkonium will differ with angle, and this will be the experimental signature of our scenario.

In the left hand plot of Fig.~\ref{fig:Angular spectra energy bin 1}, we show the contribution of the different energy bins to the angular recoil spectra of the darkonium.  We subdivide the total energy region [5, 40] keV into two separate bins: [5, 25] keV and [25, 40] keV.  From the plot we see that the majority of the angular recoil events comes from the lower nuclear recoil energy bin.  

A flattening of the angular recoil spectrum is seen for $\theta_{v_E q} \gtrsim$ 130$^\circ$ for the nuclear recoil energy bin [5, 25] keV in the left hand plot of Fig.~\ref{fig:Angular spectra energy bin 1}.  An intuitive way to understand this behavior comes from the expression for $v_{\rm min_{2}}$ in eqn.~\ref{eq:darkonium elastic Galactic frame final expression}.  For these large values of the angle $\theta_{v_E q}$ and small values of the recoil momentum $q$, there is a partial cancellation between the terms $v_E$ cos$\theta_{v_E q}$ and $q/2\mu_2$, and this explains this flattening behavior.

We try to make a simplistic estimate of the exposure required to differentiate between the darkonium scattering and 200 GeV particle scattering for this chosen normalisation.  Let us denote the number of 200 GeV particle and darkonium scattering events integrated over the full solid angle as N$_{2d}$ and N$_{d_2}$, respectively.  An exposure of 15 kg-year is required to have (N$_{d_2}$ - N$_{2d}$)$^2$/N$_{2d} \approx$ 3.  The inclusion of the experimental angular resolution, energy resolution, and other experimental uncertainties will deteriorate this ratio, and hence a larger exposure will be required to discriminate between the darkonium scattering and 200 GeV particle scattering signals.  However, our theoretical estimate shows that a reasonable exposure can distinguish between the darkonium scattering and 200 GeV particle scattering signals.

The angular distribution when the dark matter particle mass is 10 GeV and has $\sigma_{\rm el}/m$ = 1 cm$^2$ g$^{-1}$ at $v = 10$ km s$^{-1}$ is shown on the right in Fig.~\ref{fig:Angular spectra 1}.  The target nucleus is $^{19}{\rm F}$.  In this case, we have integrated over 5 keV to 14 keV to plot the angular recoil spectrum for the 10 GeV and 20 GeV dark matter particle scattering, and darkonium scattering.  

\begin{figure*}[!thpb]
\centering
\includegraphics[angle=0.0,width=0.43\textwidth]{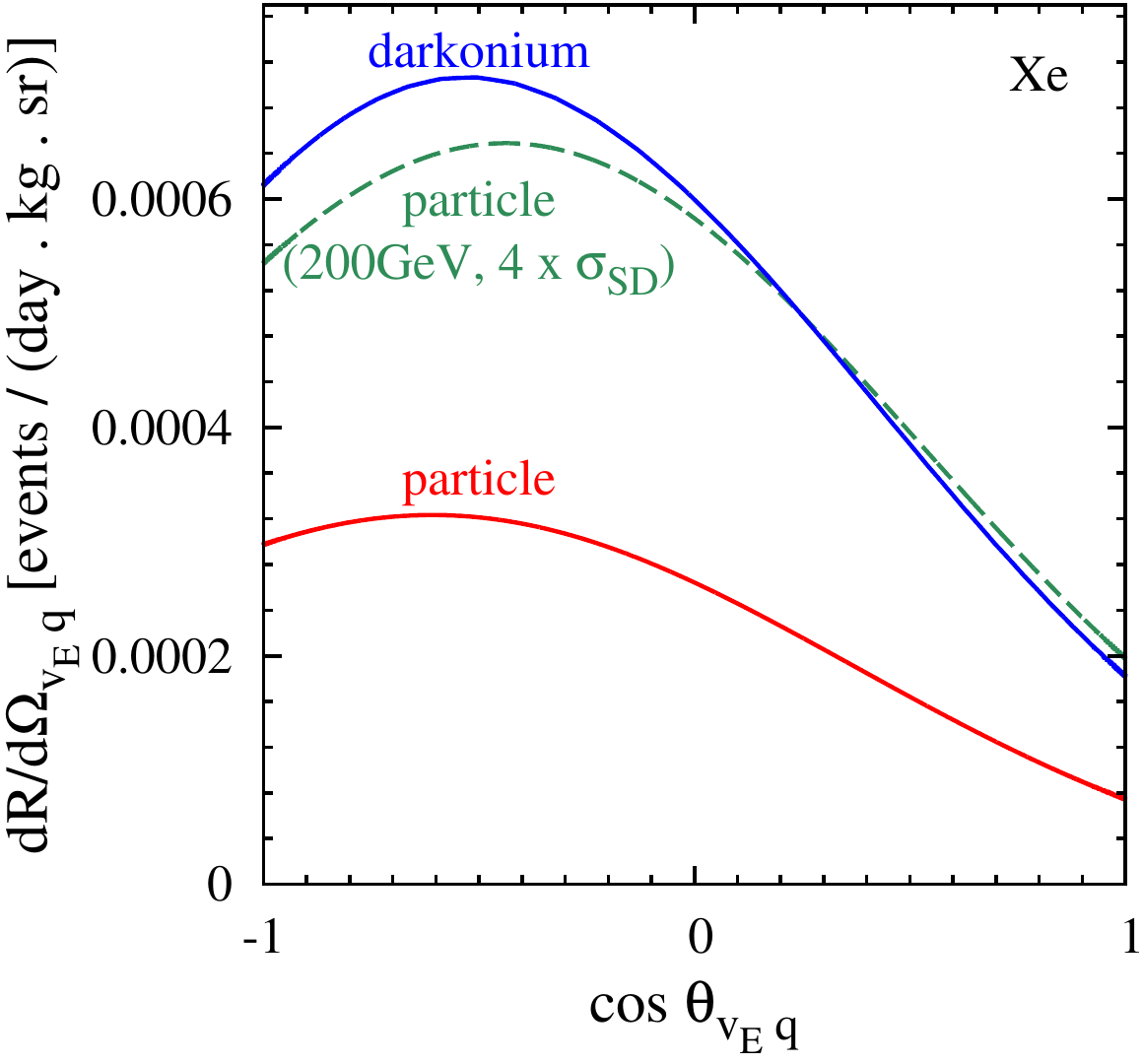}
\includegraphics[angle=0.0,width=0.43\textwidth]{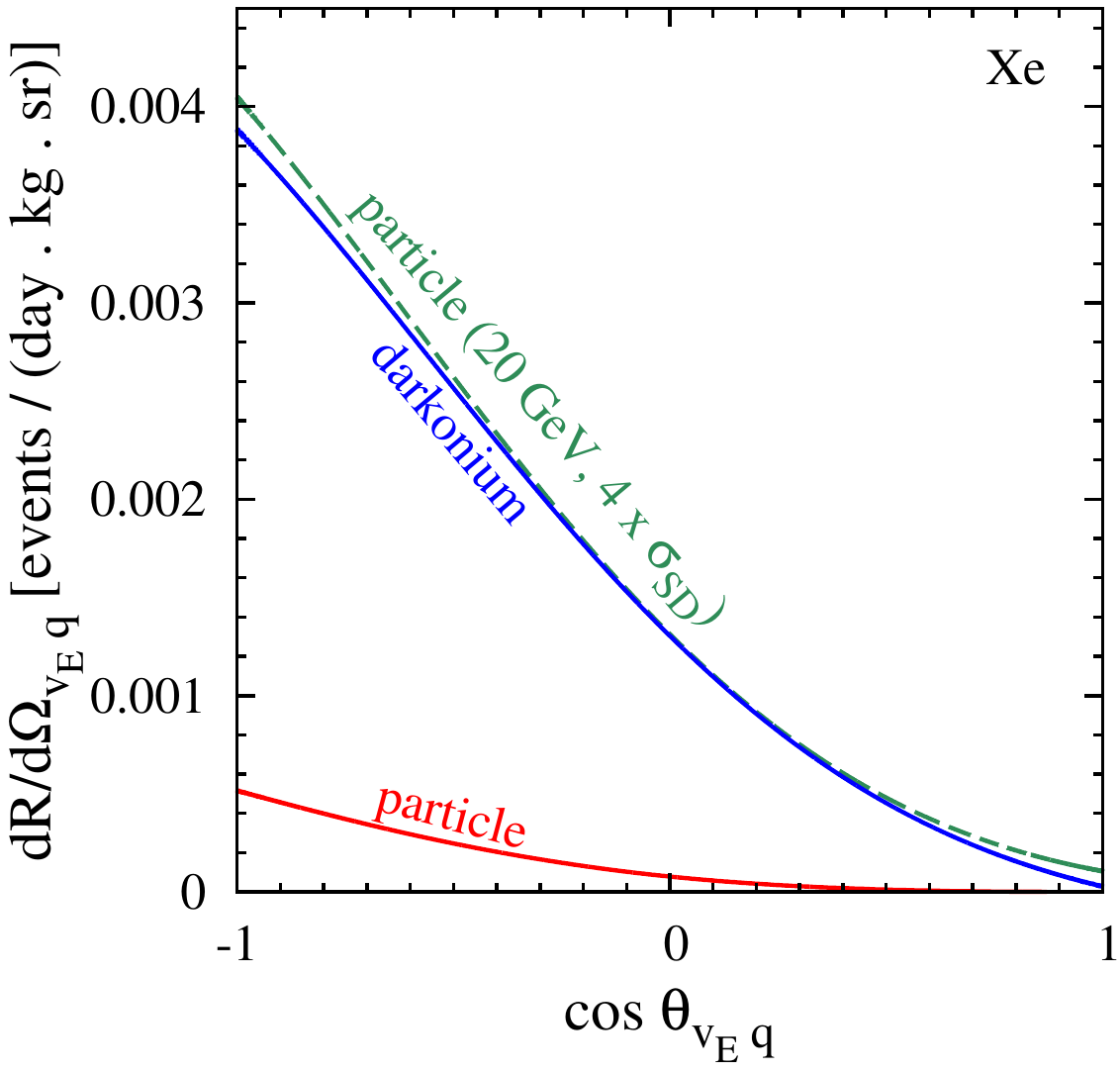}
\caption{Same as Fig.~\ref{fig:Angular spectra 1} but with xenon as the target nucleus.  {\bf Left plot:} The energy bin for integration is taken to be [2, 40] keV.  {\bf Right plot:} The energy bin for integration is taken to be [2, 23] keV.}
\label{fig:Angular spectra 2}
\end{figure*}

\begin{figure*}[!thpb]
\centering
\includegraphics[angle=0.0,width=0.43\textwidth]{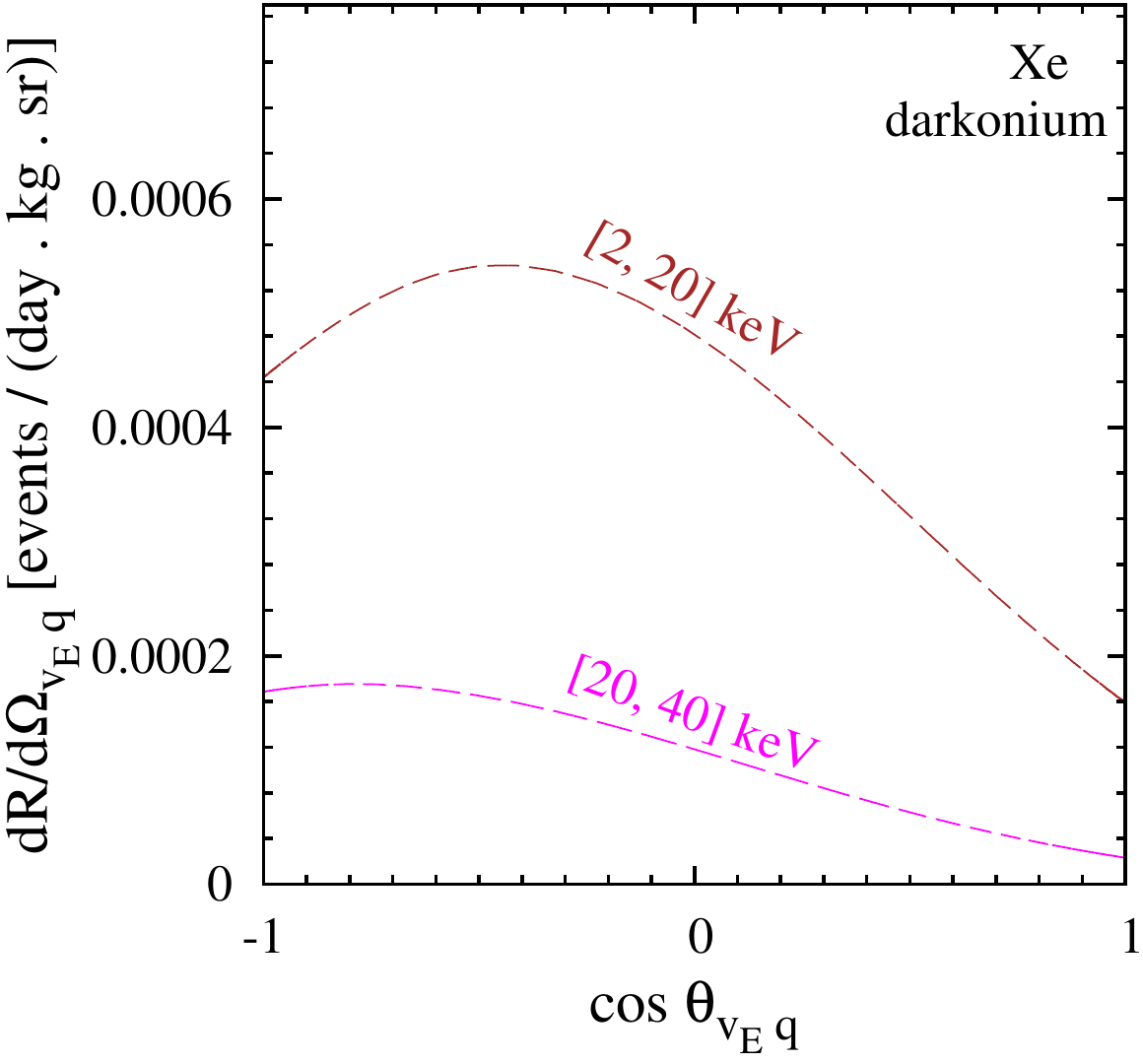}
\includegraphics[angle=0.0,width=0.43\textwidth]{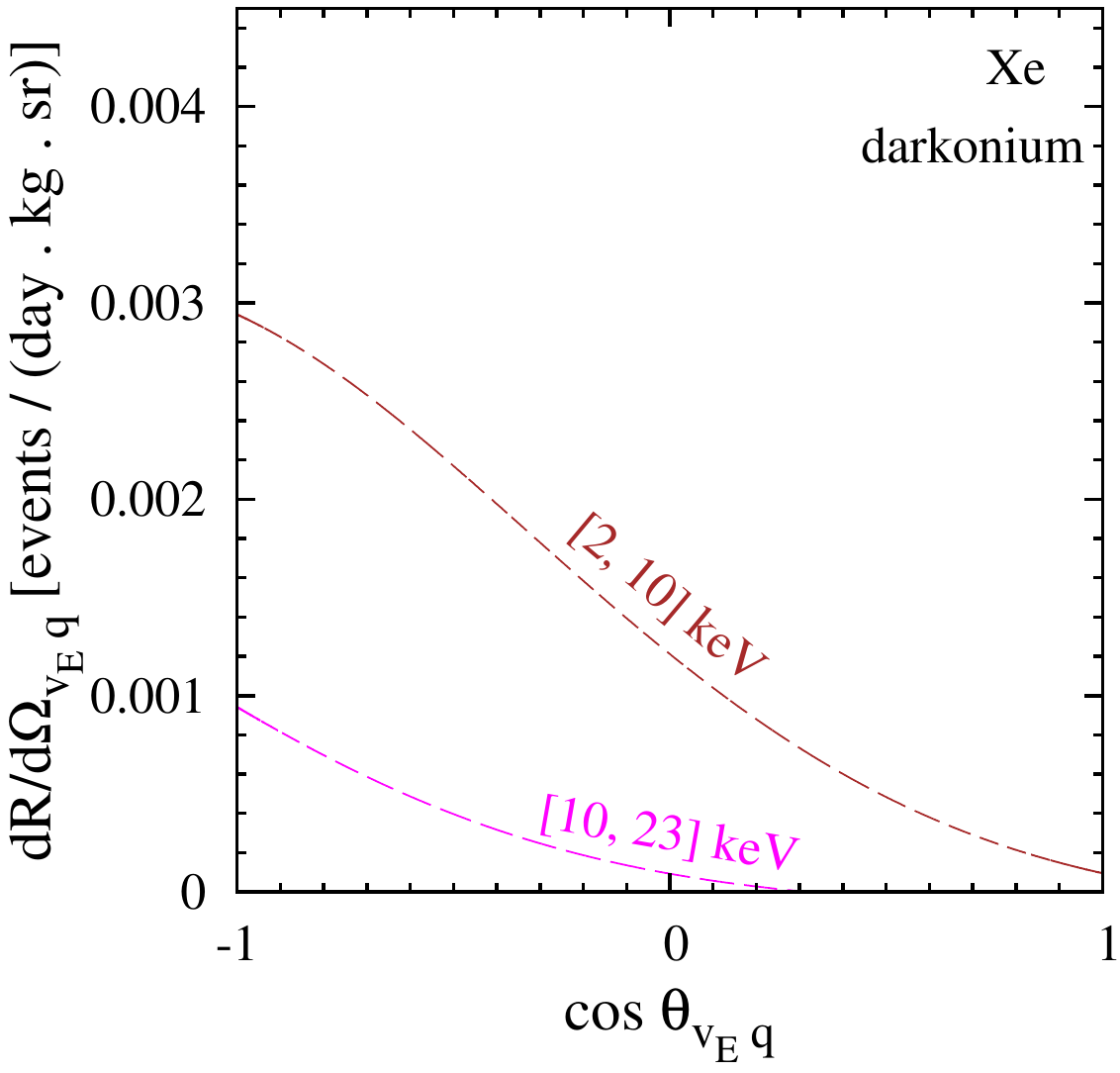}
\caption{Same as Fig.~\ref{fig:Angular spectra energy bin 1} but with xenon as the target nucleus.  {\bf Left plot:} The energy bins taken in this analysis are [2, 20] keV and [20, 40] keV when the darkonium is composed of two 100 GeV dark matter particles.  {\bf Right plot:}  The energy bins taken in this analysis are [2, 10] keV and [10, 23] keV when the darkonium is composed of two 10 GeV dark matter particles. }
\label{fig:Angular spectra energy bin 2}
\end{figure*}

\begin{figure}
\includegraphics[angle=0.0,width=0.45\textwidth]{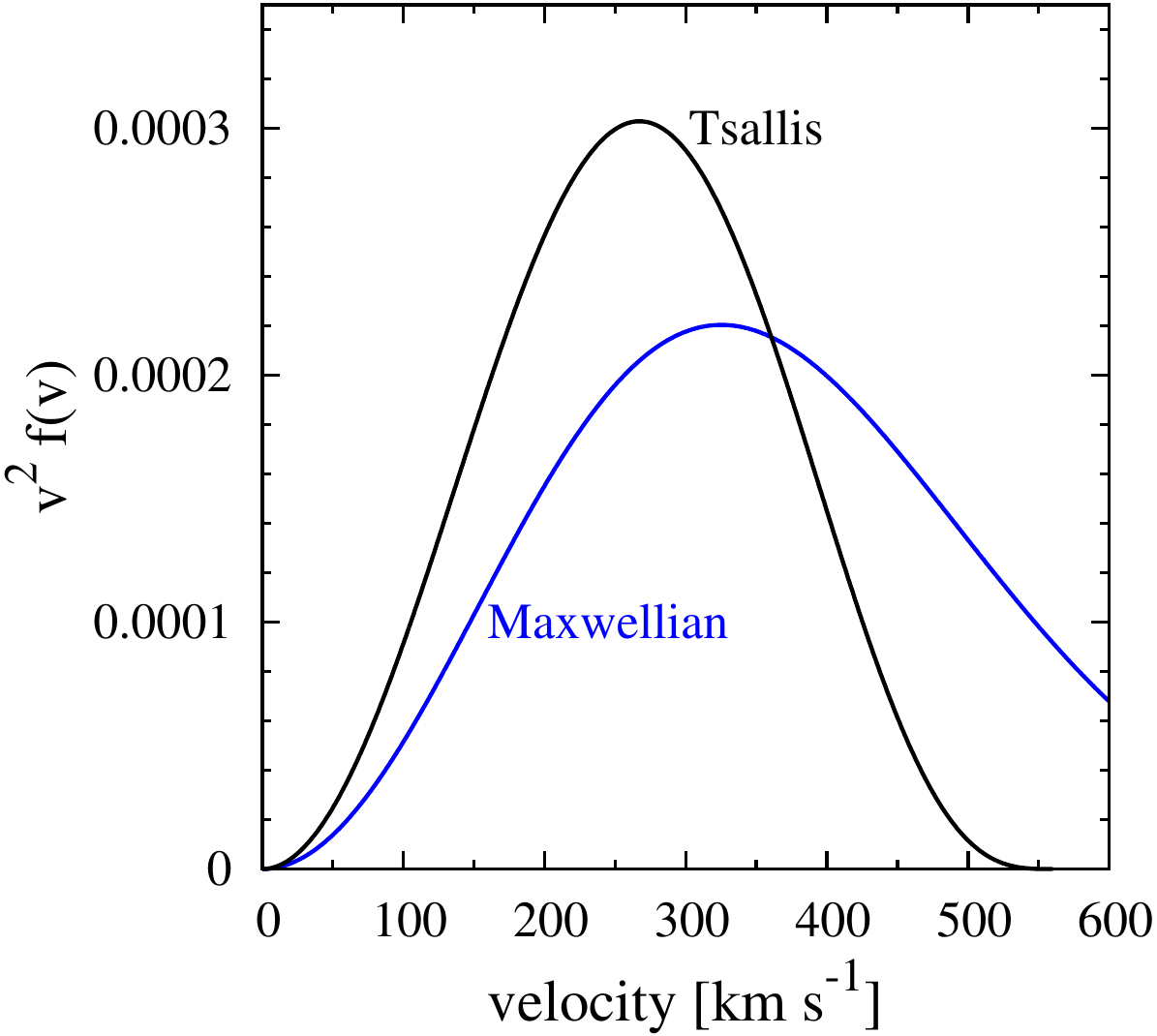}
\caption{The two different dark matter velocity profiles that are used in this work.  The Maxwellian distribution is plotted in blue.  The Tsallis distribution is plotted in black.}
\label{fig:Velocity profile}
\end{figure}

\begin{figure}
\centering
\includegraphics[angle=0.0,width=0.43\textwidth]{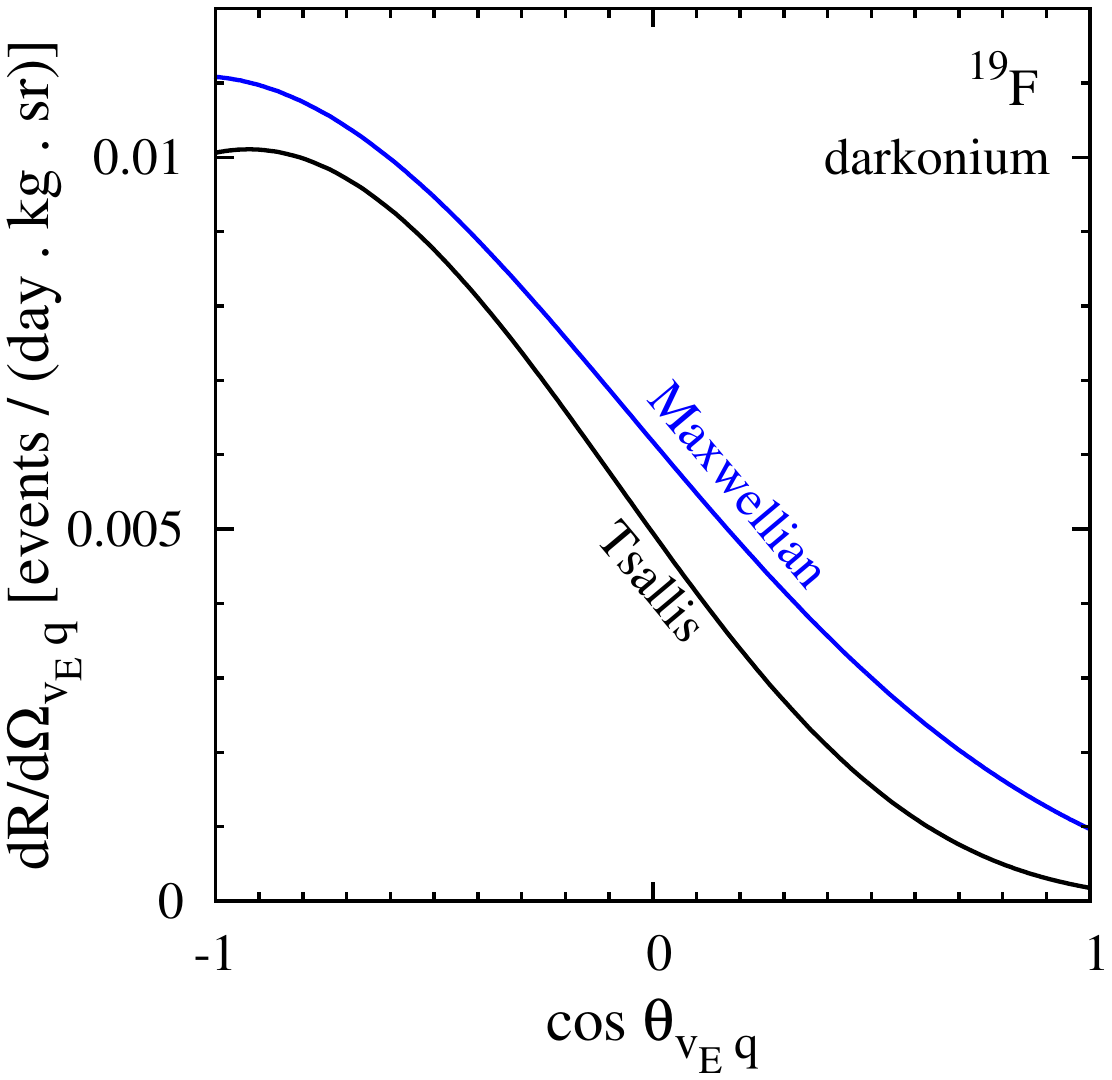}
\includegraphics[angle=0.0,width=0.43\textwidth]{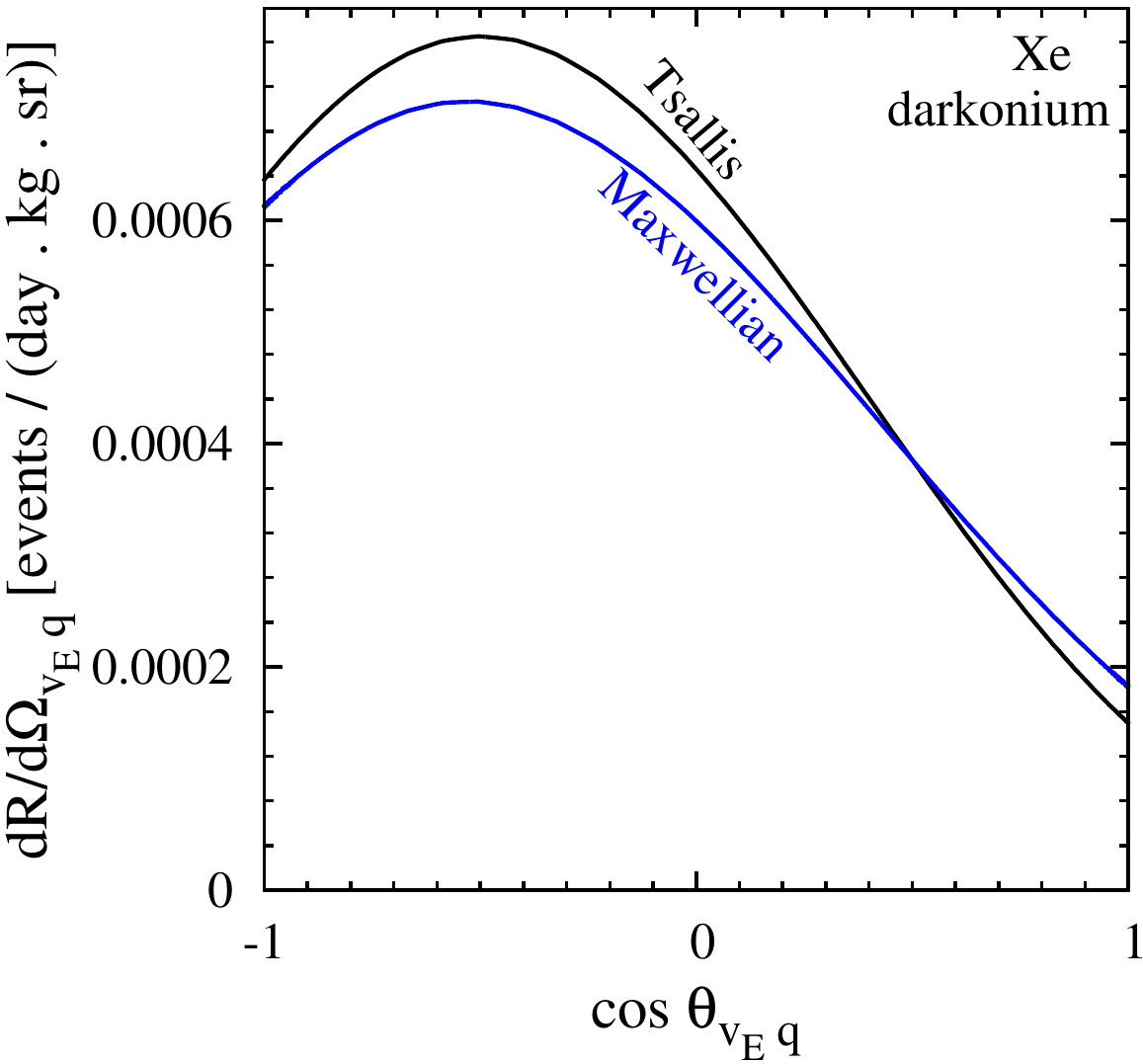}
\caption{The angular nuclear recoil spectra for darkonium scattering.  The S-wave scattering length and the normalization of the plots are the same as in the corresponding Figs.~\ref{fig:Angular spectra 1} and \ref{fig:Angular spectra 2}.  The angular recoil spectrum due to the Maxwellian velocity distribution and Tsallis velocity distribution is shown in blue and black respectively.  The darkonium is composed of two 100 GeV particles in both the plots.  {\bf Top plot :} The target is $^{19}$F.  We integrate over the nuclear energy range [5, 30] keV and [5, 40] keV to obtain this angular recoil spectrum for the Tsallis and Maxwellian distributions respectively.  {\bf Bottom plot :} The target is Xe.  We only consider $^{129}$Xe and $^{131}$Xe since we show the plot for spin-dependent interactions.  The energy bin for integration is [2, 40] keV. Note the different scales in the y-axis.}
\label{fig:Comparison of recoil spectra of MB and Tsallis}
\end{figure}

The angular recoil spectrum for the darkonium is different in both shape and normalization compared to the dark matter particle of mass 10 GeV.  The angular recoil spectrum of darkonium looks  similar to the case of a 20 GeV dark matter particle elastic scattering with a 4 times larger cross section.   This is expected as the darkonium does not break up during its collision with the nucleus.

In the right hand plot of Fig.~\ref{fig:Angular spectra energy bin 1}, we show the angular recoil spectrum when we consider a higher energy bin and a lower energy bin.  We subdivide the total recoil energy bin,  [5, 14] keV, into two parts: [5, 10] keV and [10, 14] keV.   We see that the lower energy bin contains the majority of the nuclear recoil events.  A hint of flattening of the angular recoil spectrum is seen for $\theta_{v_E q} \gtrsim$ 160$^\circ$.  Again this can be explained by the expression for $v_{\rm min_{2}}$ in eqn.~\ref{eq:darkonium elastic Galactic frame final expression}.  The reasoning is similar to the one  that explains the flattening in the lower energy bin in the left hand panel of Fig.~\ref{fig:Angular spectra energy bin 1}.  In this case, the flattening occurs at higher angles due to lower mass of the darkonium as compared to the left hand panel of Fig.~\ref{fig:Angular spectra energy bin 1}.

Denoting the number of 20 GeV particle scattering and darkonium scattering as N$_{2d, 20}$ and N$_{d_2,10}$, we find that an exposure of 5 kg-year is required for (N$_{d_2,10}$ - N$_{2d,20}$)$^2$/N$_{2d,20} \approx$ 3.  A smaller amount of exposure is required compared to the 100 GeV case due to the larger overall normalization involved.

\subsubsection{Target: Xenon}

The angular recoil spectrum when the target is Xe is shown in Fig.~\ref{fig:Angular spectra 2}.  We have taken the natural abundance of $^{129}$Xe and $^{131}$Xe as targets in 1 kg of Xe for these plots.  For the left figure, we have taken the energy bin to be 2 keV to 40 keV.  The energy bin used in the right column is 2 keV to 23 keV.  The energy threshold is again taken from Ref.~\cite{Grothaus:2014hja}.

Similar to the case where the target was $^{19}$F, the angular recoil spectrum of the darkonium is very different from that of a particle with half the mass.  As expected, for the case of the 10 GeV dark matter particle, the angular recoil spectrum of darkonium is similar to that of a 20 GeV dark matter particle with 4 times larger cross section.

We show the angular recoil spectrum for different nuclear recoil energy bins for darkonium scattering in Fig.~\ref{fig:Angular spectra energy bin 2}.  When the darkonium is composed of two 100 GeV dark matter particles, we divide the total nuclear recoil spectrum [2, 40] keV into two bins: [2, 20] keV and [20, 40] keV.   When the darkonium is composed of two 10 GeV dark matter particles, we divide the total nuclear recoil spectrum [2, 23] keV into two bins: [2, 10] keV and [10, 23] keV.  Due to the heavier mass of the target, most of the nuclear recoil events are in the lower energy bin.

A bump-like feature is seen at around 120$^{\circ}$ in the angular recoil spectrum in the left panel of Fig.~\ref{fig:Angular spectra 2}.  This feature is present for 100 GeV  and 200 GeV dark matter particle scattering and for darkonium scattering.  This feature arises due to the low threshold energy and the dark matter mass considered.  When cos\,$\theta_{v_Eq}$  is greater than $\pi$/2, the expression $v_E$ cos\,$\theta_{v_Eq}$ becomes negative, and this causes a partial cancellation between the terms in $v_{\rm min}$ in the exponential of Eqn.~\ref{eq:dark matter elastic Galactic frame final expression}.  If we take the threshold energy to be higher, say for e.g., 20 keV, then the values of $q/2\mu$ are sufficiently large enough to not cause a cancellation with the term $v_E$ cos\,$\theta_{v_Eq}$ and the bump-like feature disappears.  A similar reasoning is also applicable for the appearance of the velocity-dependent part feature in the case of darkonium scattering.

\subsection{Impact of non-Maxwellian velocity distribution}
\label{sec:non-Maxwellian}

Numerical simulations of Milky Way like object which includes dark matter and baryons often predict a velocity distribution which is very different from the Maxwellian distribution that we have assumed till now.  In particular, Ref.~\cite{Ling:2009eh} finds that the velocity distribution closely follows the Tsallis distribution: 
\begin{eqnarray}
f(v) = N_{\rm Tsallis} \bigg \{ 1 - (1-q) \dfrac{v^2}{v_0^2} \bigg\}^{\dfrac{q}{1-q}}
\label{eq:Tsallis distribution}
\end{eqnarray}
where $v_0$ = 267.2 km s$^{-1}$ and $q = 0.773$.  The normalization constant $N_{\rm Tsallis}$ is obtained from $\int d^3 v \, f(v) = 1$.  The Tsallis distribution shows that the maximum velocity of dark matter particles is $v_{\rm max} = \{ v_0^2/(1-q) \}^{1/2} \approx$ 560 km s$^{-1}$.

We compare these different velocity profiles in Fig.~\ref{fig:Velocity profile}.  The $v^2 \, f(v)$ of the Tsallis distribution peaks at around 250\,km s$^{-1}$, whereas it peaks at around 300 km s$^{-1}$ for the Maxwellian distribution.  This is not the only type of non-Maxwellian velocity distribution seen in simulations of Milky Way-like galaxies which include baryons.  Non-Maxwellian velocity profile is also seen in more modern simulations of the Milky Way which includes baryons~\cite{Butsky:2015pya}.

The inclusion of this non-Maxwellian velocity distribution in our calculation is straight forward.  Closed form expression of the nuclear recoil energy distribution is not possible for this Tsallis distribution.  For the elastic scattering of a dark matter particle with a nucleus, we use the expression of $f(v)$ in Eqn.~\ref{eq:Tsallis distribution} in Eqn.~\ref{eq:eq:dark matter elastic Galactic frame integrated over theta_vq}.  The integration over $v$ can be carried out numerically.  The non-Maxwellian expression of $f(v)$ is used in Eqn.~\ref{eq:differential sigmav darkonium elastic scattering}.  Similarly the non-Maxwellian expression for $f(v)$ is used in Eqn.~\ref{eq:darkonium break up scattering rate in the laboratory}.

We compare the angular recoil spectrum for the Maxwellian velocity distribution and Tsallis velocity distribution in Fig.~\ref{fig:Velocity profile}.  It is clear from the figures that the shape of the angular recoil energy spectrum is sufficiently different for the two different angular recoil spectra.

In the top panel of Fig.~\ref{fig:Velocity profile}, we plot the angular nuclear recoil spectra when a darkonium, composed of two 100 GeV dark matter particles, collides with $^{19}$F for the two different velocity distributions that we consider, Maxwellian and Tsallis.  Due to the lower maximum velocity in the Tsallis distribution, the integration range is [5, 30] keV.  This also explains why the angular recoil distribution due to the Tsallis distribution is lower than the one due to the Maxwellian distribution.  The values of the S-wave scattering length and the dark matter - nucleon cross section that we consider in this plot is the same as in Fig.~\ref{fig:Angular spectra 1}.

In the bottom panel of Fig.~\ref{fig:Velocity profile}, we plot the angular nuclear recoil spectra when a darkonium, composed of two 100 GeV dark matter particles, collides with Xe for the two different velocity distributions that we consider, Maxwellian and Tsallis.  We consider spin-dependent cross sections with the same parameters as in Fig.~\ref{fig:Angular spectra 1}.  Again, the shape of the angular nuclear recoil spectra is different for the Maxwellian and Tsallis spectra.

If the smoking gun signature for a darkonium is observed in a dark matter directional detection experiment, then the shape of this angular nuclear recoil spectrum can be used to reconstruct the underlying dark matter velocity distribution.  These new angular recoil spectra as derived in this work open up a new avenue to probe exotic properties of dark matter like strong self-interaction.  Although the normalization in all our plots is arbitrary, the shape of the angular recoil spectrum is uniquely dictated by the S-wave scattering length.  In this work, we derived the angular recoil spectrum due to a specific value of the self-interaction cross section and for two specific dark matter velocity distributions.  Variations on this theme require considering different values of the self-interaction cross section as advocated in Ref.~\cite{Elbert:2014bma} and considering various different dark matter velocity distributions~\cite{Chaudhury:2010hj,Kuhlen:2009vh,Bhattacharjee:2012xm,Bhattacharjee:2013exa,Mao:2012hf,Mao:2013nda}.

\section{Conclusion}
\label{sec:conclusion}

We have discussed the directional detection signal that is expected when a bound-state dark matter collides with a nucleus.  The bound state in our case is motivated by the hints of strong self-interaction cross section between dark matter particles.  The predictive assumption of a near threshold S-wave resonance is used to uniquely determine the angular recoil spectrum.

The S-wave scattering length determines the self-interaction cross section between the dark matter particles and also determines the binding energy of the resultant bound state (which we call darkonium).  When the darkonium is incident on a nucleus, two possibilities arise: ($i$) the darkonium elastically scatters with the nucleus, such that the angular recoil spectrum contains information about the form factor of the darkonium, which is uniquely determined by the S-wave scattering length, and  ($ii$) the darkonium breaks up while scattering with the nucleus.  Even in the latter case the angular recoil spectrum is uniquely determined by the S-wave scattering length.  

The angular recoil spectrum for two different targets and two different dark matter masses are shown in Figs.~\ref{fig:Angular spectra 1} and \ref{fig:Angular spectra 2}.  Figs.~\ref{fig:Angular spectra energy bin 1} and \ref{fig:Angular spectra energy bin 2} show the angular recoil spectrum when divided into different energy bins.  We take $\sigma_{\rm el}/m$ = 1 cm$^2$ g$^{-1}$ at relative velocity $v$ = 10 km s$^{-1}$ to determine the S-wave scattering length in all the cases.  For the case of the 10 GeV dark matter particle mass, the bound state does not break up during its collision with the nucleus.  In this case, the angular recoil spectrum of the incident darkonium is very similar to that of a dark matter particle of mass 20 GeV.  When the dark matter particle mass is 100 GeV, the angular recoil spectrum of the dakonium is different from the angular recoil spectrum of either 100 GeV or 200 GeV dark matter particle mass.

Figs.~\ref{fig:Angular spectra 1} to \ref{fig:Angular spectra energy bin 2} assume that the underlying dark matter velocity distribution is Maxwellian.  Simulations of Milky Way-sized halos which include baryons typically predict a non-Maxwellian dark matter velocity distribution, for e.g., Ref~\cite{Ling:2009eh} and \cite{Butsky:2015pya}.  We compare the Maxwellian and Tsallis distribution in Fig.~\ref{fig:Velocity profile}.  Fig.~\ref{fig:Comparison of recoil spectra of MB and Tsallis} compares the angular nuclear recoil spectrum when a darkonium scatters with $^{19}$F and Xe nuclei for a Maxwellian and Tsallis dark matter velocity distribution.  As expected the angular recoil spectrum is  different for different dark matter velocity distributions.

The predictive nature of the underlying physics implies that if these signatures are detected in a future dark matter directional detection experiment, then many of the low-energy properties of the dark matter will be completely determined.  Such a smoking gun signature from a model-independent approach will be crucial in determining the underlying particle properties of dark matter.

\section*{Acknowledgments} 

We thank John Beacom, Eric Braaten, Debtosh Chowdhury, Basudeb Dasgupta, Bhaskar Dutta, Yu Gao, Daekyoung Kang, Matthew Kistler, Rafael Lang, Philipp Mertsch, Kenny C.Y. Ng, Annika Peter, Carsten Rott, Stephen Sadler, and Louis Strigari for discussions.  This work was supported by KIPAC.  We also received partial support from MIAPP of the DFG cluster of excellence ``Origin and Structure of the Universe".

\bibliographystyle{kp}
\interlinepenalty=10000
\tolerance=100
\bibliography{Bibliography/references}

\end{document}